\documentclass[12pt,preprint]{aastex}
\usepackage{natbib}
\def\be{\begin{equation}}
\def\ee{\end{equation}}

\def\m{~$\mu$m}
\def\R25{$R_{25}$}

\def\HI   {\ion{H}{1}}

\begin {document}

\title{RADIAL STAR FORMATION HISTORIES IN FIFTEEN NEARBY GALAXIES}
\shorttitle{Age Gradients}

\author {
Daniel~A. Dale\altaffilmark{1},
Gillian~D. Beltz-Mohrmann\altaffilmark{2},
Arika~A. Egan\altaffilmark{3},
Alan~J. Hatlestad\altaffilmark{1},
Laura~J. Herzog\altaffilmark{4},
Andrew~S. Leung\altaffilmark{5},
Jacob~N. McLane\altaffilmark{5},
Christopher Phenicie\altaffilmark{6},
Jareth~S. Roberts\altaffilmark{1},
Kate~L. Barnes\altaffilmark{7},
M\'ed\'eric Boquien\altaffilmark{8},
Daniela Calzetti\altaffilmark{9},
David~O. Cook\altaffilmark{1},
Henry~A. Kobulnicky\altaffilmark{1}, 
Shawn~M. Staudaher\altaffilmark{1}, and
Liese van Zee\altaffilmark{7}
}
\altaffiltext{1}{Department of Physics \& Astronomy, University of Wyoming, Laramie WY; ddale@uwyo.edu}
\altaffiltext{2}{Department of Physics, Wellesley College, Wellesley MA}
\altaffiltext{3}{Department of Physics, Northern Michigan University, Marquette MI}
\altaffiltext{4}{Department of Physics \& Astronomy, Minnesota State University, Moorhead MN}
\altaffiltext{5}{Department of Astronomy, University of Texas, Austin TX}
\altaffiltext{6}{School of Physics \& Astronomy, University of Minnesota, Minneapolis MN}
\altaffiltext{7}{Department of Astronomy, Indiana University, Bloomington IN}
\altaffiltext{8}{Unidad de Astronom\'ia, Universidad de Antofagasta, Antofagasta, Chile}
\altaffiltext{9}{Department of Astronomy, University of Massachusetts, Amherst MA}

\begin {abstract}
New deep optical and near-infrared imaging is combined with archival ultraviolet and infrared data for fifteen nearby galaxies mapped in the {\it Spitzer} Extended Disk Galaxy Exploration Science survey.  These images are particularly deep and thus excellent for studying the low surface brightness outskirts of these disk-dominated galaxies with stellar masses ranging between $10^8$ and $10^{11}~M_\odot$.  The spectral energy distributions derived from this dataset are modeled to investigate the radial variations in the galaxy colors and star formation histories.  Taken as a whole, the sample shows bluer and younger stars for larger radii until reversing near the optical radius, whereafter the trend is for redder and older stars for larger galacto-centric distances.  These results are consistent with an inside-out disk formation scenario coupled with an old stellar outer disk population formed through radial migration and/or the cumulative history of minor mergers and accretions of satellite dwarf galaxies.  However, these trends are quite modest and the variation from galaxy to galaxy is substantial.  Additional data for a larger sample of galaxies are needed to confirm or dismiss these modest sample-wide trends.
\end {abstract}

\section{Introduction}
\label{sec:intro}

Understanding how galaxies assemble is crucial to understanding the overall picture of galaxy evolution.  In $\Lambda$CDM models, galactic disks are built through mergers and the accretion of small satellites, as well as through {\it in situ} star formation activity \citep[e.g.,][]{abadi03,governato04,robertson06,governato07}.  In the ``inside-out'' scenario, galaxy disks systematically grow through star formation in the outermost parts.  This process should manifest itself with specific observable hallmarks easily obtained from ground-based campaigns, including gradients in metallicity and specific star formation rates coupled with longer disk scale lengths for younger stellar populations and thus increasingly blue colors at larger radii \citep{larson76,ryder94,avilareese00,dejong96,macarthur04,munozmateos07,wang11,barnes14}.  Resolved stellar photometry can also help provide clues to the formation history of spiral galaxies \citep[e.g.,][]{brown08,monachesi13}.  Analysis of star formation histories via {\it Hubble Space Telescope}-based color-magnitude diagrams for NGC~300 and M~33 indicate an inside-out formation scenario, with decreasing metallicity and stellar age for larger radii within the galaxy disks, and then a reversal to positive stellar age gradients for the outer portions of M~33 \citep{williams09,gogarten10,barker11}.  Galaxy-wide outside-in disk formation, on the other hand, has only been plausibly suggested for dwarf galaxies, galaxies for which their low masses leave them more susceptible to environmental effects \citep[e.g.,][]{zhang12}.  

Investigations that concentrate on the outermost parts of galaxies can yield additional information on the galaxy formation process \citep[e.g.,][]{ferguson98,thilker07}.  For example, \cite{pohlen06} find that later-type spiral galaxies are more likely to exhibit downbending, where the outer surface brightness profile more steeply drops than the surface brightness profile of the main disk, and earlier-type spirals are more likely to have outer upbending profiles.  Simulations predict this upbending phenomenon in galaxies with significant accretion histories \citep{abadi06}.  The Keck-based work of \cite{ibata05} finds stars kinematically associated with M~31 as far as 70~kpc (1.6$a_{25}$\footnote{$a_{25}$ is defined as the length of the semi-major axis for the $B$-band isophote at 25~mag~arcsec$^{-2}$ \citep{devaucouleurs91}.}) from the galaxy center, located in complex, ephemeral substructures suggestive of prior accretion.  Similarly, \cite{herbertfort09} use deep LBT optical and GALEX ultraviolet imaging to probe the blue and red stellar populations in NGC~3184 out to and beyond 1.6$a_{25}$.  In this galaxy red stellar clusters are more prominent than their blue counterparts at the largest radii; the authors posit that the the stellar clusters at the outer edge of the main stellar disk are a natural extension of the spiral arms whereas the outermost stellar clusters likely arrived via past accretion events \citep[see also][]{herbertfort12}.  The presence of older stars in a galaxy's periphery could arise through internal dynamical processes that redistribute stars originally formed near the center \citep[e.g., ``radial migration''][]{sellwood02,roskar08,radburnsmith12}.  Alternatively, simulations suggest that the stellar haloes of galaxies arise from the cumulative history of hierarchical merging and very little stemming from in situ star formation \citep{abadi06,read06,purcell07,cooper13}.  If these accretion events are largely ancient \citep[e.g.,][]{bullock05,abadi06} then passive stellar evolution will naturally result in predominantly red stellar outskirts.  Though it is challenging to disentangle which of the various processes are the main drivers in building up galaxy disks and haloes, a necessary first step is to construct a robust, wide-field, and multi-wavelength dataset for constraining galaxy surface brightness profiles, extinctions, metallicities, and overall star formation histories.

Here we report on a multi-wavelength analysis of 15 nearby galaxies drawn from the Extended Disk Galaxy Exploration Science (EDGES) survey \citep{vanzee12}.  One of the primary goals of EDGES is to probe the extent of the old stellar population as far out as possible with {\it Spitzer}, and to compare those extents at other wavelengths tracing other emission mechanisms.  In this effort we combine the 3.6\m\ EDGES data with new deep, ground-based optical and archival space-based ultraviolet and infrared imaging to constrain the radial trends in the star formation histories of these 15 galaxies.  The depth of our imaging dataset enables a careful investigation in particular of the galaxy outskirts, out to 1.5 times the de Vaucouleurs radius, where the low surface brightness levels typically are a challenge to detect with sufficient signal-to-noise.  The mid- and far-infrared data utilized in this analysis allow the SED fits to be carried out in an energy-balanced fashion: any ultraviolet/optical light that is modeled to be extinguished by dust reappears at longer wavelengths as dust emission.  Previous EDGES work includes \cite{barnes14}, who utilize 3.6\m, \HI, and far-ultraviolet data to characterize the full disk and associated streamer for NGC~5236 (M~83), \cite{richards15} combine \HI\ and 3.6\m\ data with optical imaging and spectroscopy to constrain a dark matter halo model for NGC~5005, and \cite{staudaher15a} use the 3.6\m\ data to quantify the stellar halo mass fraction of NGC~5055 (M~63) and the mass of its prominent tidal feature.  A detailed study of the 3.6\m\ surface brightness profiles will be presented in \cite{staudaher15b}.

Section~\ref{sec:sample} presents the sample studied here, Section~\ref{sec:data} reviews the new and archival data compiled for this analysis along with an overview of the data processing, Section~\ref{sec:analysis} explains the analysis including the fitting of the spectral energy distributions (SEDs), Section~\ref{sec:results} presents the main results, and Section~\ref{sec:summary} provides a summary and brief discussion.

\section{Galaxy Sample}
\label{sec:sample}

Table~\ref{tab:sample} provides the list of galaxies studied here.  They represent a subset of EDGES galaxies observable during the summer of 2014 from the Wyoming Infrared Observatory (WIRO) and served as the centerpiece of the work carried out by the 2014 cohort of Wyoming REU (Research Experience for Undergraduates) interns.  Priority was given to targets with good ancillary ultraviolet and infrared data (\S~\ref{sec:data}).  The overall EDGES sample contains 92 nearby galaxies spanning a range of morphology, luminosity, and environment, for galaxies at high Galactic latitudes $|b|>60$\degr, with apparent magnitudes $m_B<16$, and optical angular diameters $2 \lesssim D(^\prime) \lesssim 13$.  Included in the list of 15 galaxies studied here is one non-EDGES ``Target of Opportunity'' galaxy, NGC~4625, which happened to fall within the field-of-view of our NGC~4618 observations.  Two of the galaxies in this subset of EDGES have S0 morphology, four are irregulars, and the remainder are spiral galaxies.  All 15 galaxies lie within $\sim 20$~Mpc.

\section{Data}
\label{sec:data}

\subsection{{\it Spitzer} 3.6\m\ Data}
For the EDGES sample we constructed large mosaics based on {\it Spitzer Space Telescope} imaging that trace 3.6 and 4.5\m\ substructures out to at least five times the optical radius $a_{25}$.  Compared to most {\it Spitzer}/IRAC imaging campaigns of nearby galaxies, the EDGES near-infrared image mosaics are quite deep; the 1800~s integration per position obtained for EDGES targets is 7.5 times longer than what was obtained for the SINGS \citep{kennicutt03}, LVL \citep{dale09b}, and S$^4$G \citep{sheth10} surveys, and 12--30 times longer compared to the IRAC GTO project \citep{pahre04}.  We reach a 1$\sigma$ per pixel sensitivity of 2~kJy~sr$^{-1}$, and averaging over several square arcminutes can reach down to below 0.4~kJy~sr$^{-1}$ (fainter than 29~mag~arcsec$^{-2}$ AB), a level necessary for securely detecting faint stellar streams associated with nearby galaxies \citep{krick11,barnes14,staudaher15a}.  For comparison, {\it WISE} achieves a 3.4\m\ diffuse sensitivity larger than 1~kJy~sr$^{-1}$ over a $5^\prime$$\times$$5^\prime$ area \citep{wright10}.  The photometric accuracy is taken to be 5\% \citep[see also][]{reach05}.

\subsection{Ancillary Ultraviolet and Infrared Data}
Archival ultraviolet and infrared data were gathered from the {\it GALEX}, {\it Spitzer}, {\it WISE}, and {\it Herschel Space Observatory} archives.  In all instances, priority was given to longer-exposure data, where available.  As such, the ultraviolet data for only two galaxies (see Table~\ref{tab:integrations}) stem exclusively from the shallow All-Sky Imaging Survey for which the integrations are $\sim$0.1~ks \citep{martin05}; the majority of the GALEX imaging utilized here arises from integrations longer than 1~ks.  For these longer integrations the limiting surface brightnesses are $\sim$32.5 and 32.0~mag~arcsec$^{-2}$ AB, respectively for the far-ultraviolet and near-ultraviolet channels, based on the standard deviations in the median sky values for a set of sky apertures placed beyond the galaxy emission (see \S~\ref{sec:analysis}); for the small set of shorter integration ultraviolet images, the surface brightness limits are about 1--2 mag~arcsec$^{-2}$ shallower.  Likewise, most of the infrared data on warm dust emission derives from {\it WISE} 12\m\ \citep{wright10} and {\it Spitzer} 24\m\ imaging \citep{dale09b}, with {\it WISE} 22\m\ and {\it Herschel} 70\m\ used to fill in any gaps in the {\it Spitzer} 24\m\ archives.  The surface brightness sensitivity of the infrared data is approximately 26.7 AB mag~arcsec$^{-2}$ at 12\m, 25.0~mag~arcsec$^{-2}$ at 22\m, 26.0~mag~arcsec$^{-2}$ at 24\m, and 19.5~mag~arcsec$^{-2}$ at 70\m\ \citep[see also][]{wright10,dale12}.  The GALEX photometric accuracy is estimated at 0.05~mag \citep{gildepaz07,morrissey07} and the infrared calibrations are known to $\sim$10\%, 10\%, 7\%, and 5\% at 12, 22, 24, and 70\m, respectively \citep{wright10,dale07,dale12}.  These ancillary/archival datasets have angular resolutions of $\sim 5-6$\arcsec.

\subsection{New Optical Observations and Data Processing}
\label{sec:new}
\begin{figure}
 \plotone{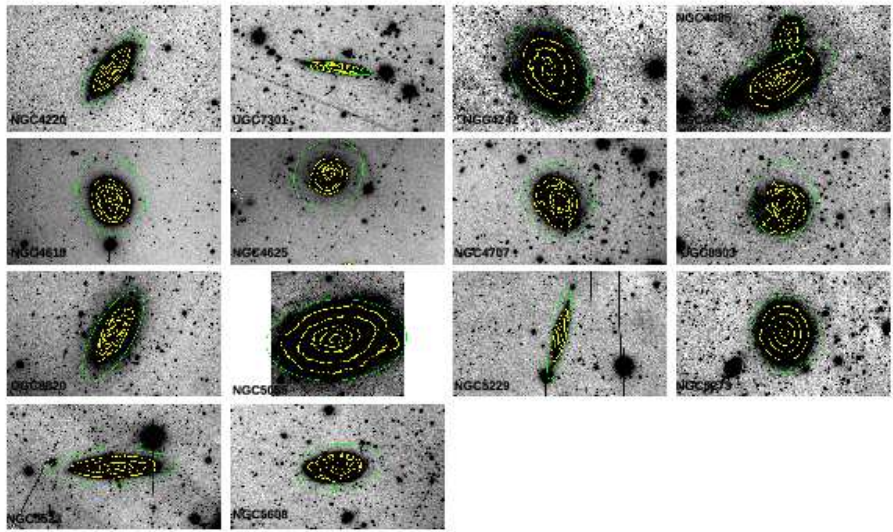}
 \caption{{\it Contact the first author or see the version in AJ for the high-res version.}  Mosaic of {\it r$^\prime$} images for the sample.  North is up, East is to the left.  The ellipses indicate the outermost-extent probed by the photometry described in \S~\ref{sec:photometry}.  Select contours are overlaid to highlight structural features.}
 \label{fig:mosaic}
\end{figure}

New deep {\it u$^\prime$g$^\prime$r$^\prime$} imaging was obtained on the WIRO 2.3~m telescope with the WIROPrime camera \citep{pierce02} over the course of the summer of 2014 (Figure~\ref{fig:mosaic}).  For each galaxy and each filter 12 individual 300~s frames were taken.  Individual frames were randomly dithered with small offsets for enhanced pixel sampling.  Each night a series of zero second bias frames were obtained in addition to a series of twilight sky flats within each filter.  

The optical images were processed with standard procedures, including subtraction of a master bias image and removal of pixel-to-pixel sensitivity variations through flatfield corrections.  Typically the sky flat was constructed from flats taken on the same night, but occasionally flats from multiple consecutive nights were utilized.  The 12 dithered 300~s frames for a galaxy taken in one filter were aligned and stacked, resulting in images with integrations equivalent to one hour.  The astrometric solutions and flux zeropoints were calibrated using positions and photometry extracted from Sloan Digital Sky Survey \citep[SDSS;][]{york00} imaging on several foreground stars spread across each image stack.  Scattered light from Galactic dust is fortunately not an issue for these observations, as all 15 targets lie north of $b=+$66$\deg$.  The uncertainties in the zeropoint calibrations were typically 3\%.  The 5$\sigma$ {\it u$^\prime$g$^\prime$r$^\prime$} point source sensitivities are $\sim$23.0, 24.0, 23.4~mag AB for 2\farcs8 diameter apertures (twice the seeing FWHM), about 1--1.8 magnitudes deeper than SDSS \citep[e.g., sdss.org/dr12/scope;][]{cook14a}.  The stacked images are flat to 1\% or better on 10\arcmin\ scales.  The limiting {\it u$^\prime$g$^\prime$r$^\prime$} surface brightnesses are $\sim$28.2, 28.4, and 28.0~mag~arcsec$^{-2}$ AB, based on the standard deviations in the median sky values for a set of sky apertures placed beyond the galaxy emission (see \S~\ref{sec:analysis}), about 1~mag~arcsec$^{-2}$ deeper than SDSS \citep{pohlen06,dsouza14}.  Similar to what was done for the imaging at all other wavelengths studied here, foreground stars and background galaxies were removed from each optical image using IRAF/{\tt IMEDIT} and a local sky interpolation.  This editing typically reached down to sources of several microJanskys ($\sim21-22$~mag AB).  Of particular note is the bright foreground star superposed on the nucleus of NGC~4707.  This star was edited in the {\it u$^\prime$g$^\prime$r$^\prime$} and 3.6\m\ images.  Additionally, the overlap region between NGC~4485 and NGC~4490 was edited.  Fortunately this region is fractionally small compared to the total areas within the outermost elliptical annuli: approximately $\frac{1}{4}$ for NGC~4485 and $\frac{1}{8}$ for NGC~4490.

\section{Data Analysis}
\label{sec:analysis}

\begin{figure}
 \plotone{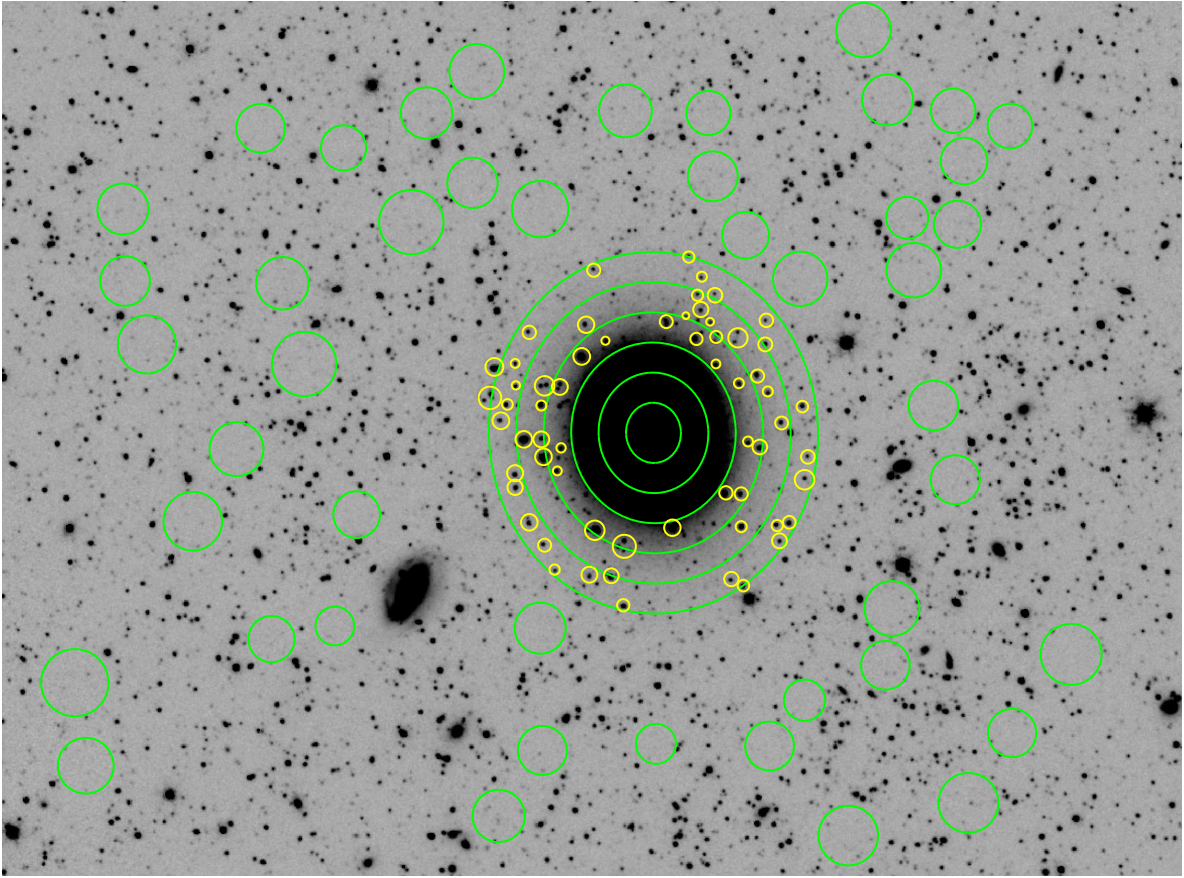}
 \caption{A 16\farcm5$\times$9\farcm7 portion of the {\it Spitzer} 3.6\m\ mosaic of NGC~5273.  The ellipses demonstrate the annular regions for extracting photometry and the large and small circles respectively show the sky apertures and foreground star masks.  North is up, East is to the left.}
 \label{fig:aps}
\end{figure}

To facilitate a consistent panchromatic analysis, the higher angular resolution data ({\it u$^\prime$g$^\prime$r$^\prime$}, 3.6\m) were smoothed (using a Gaussian smoothing profile) to the $\sim$6\arcsec\ resolution of the ultraviolet and mid-infrared data.  However, an identical analysis using the images at their native resolutions yields similar end results described in \S~\ref{sec:results}, due in large part to the relatively coarse annular apertures utilized (see \S~\ref{sec:photometry}).  A local sky value was estimated and removed via a set of apertures located just beyond the outermost reaches of the galaxy emission (Figure~\ref{fig:aps}).

\subsection{Elliptical Photometry}
\label{sec:photometry}

Photometry was obtained for each galaxy using IRAF/{\tt IMCNTS} and a series of elliptical annuli covering semi-major axis $a$ ranges of
0$<$${a} \over {a_{25}}$$<$0.25, 0.25$<$${a} \over {a_{25}}$$<$0.5, 0.5$<$${a} \over {a_{25}}$$<$0.75, 0.75$<$${a} \over {a_{25}}$$<$1, 1$<$${a} \over {a_{25}}$$<$1.25, and 1.25$<$${a} \over {a_{25}}$$<$1.5.
For the range of galaxy angular sizes in our sample, these $\frac{1}{4} a_{25}$ annular widths correspond to a range of values between 14\arcsec\ (for UGC~7301) and 94\arcsec\ (for NGC~5055).  Traditional surface brightness analyses typically rely on a finer set of annular widths, but any radial analysis that requires a full panchromatic dataset (e.g., SED fitting) is necessarily limited by the wavelength for which the detected extent of the emission is smallest.  Our choice of coarsely-spaced annular apertures provides less spatial information but allows for more robust signal-to-noise in the galaxy outskirts where the emission is weakest.  All annuli for a given galaxy used the same (NED-based) centroids, position angles, and ellipticities.  Photometric uncertainties $\epsilon_{\rm total}$ are computed by summing in quadrature the calibration error $\epsilon_{\rm cal}$ and two uncertainties $\epsilon_{\rm sky,local}$ and $\epsilon_{\rm sky,global}$ based on the measured sky fluctuations, i.e., 
\be
\epsilon_{\rm total} = \sqrt{\epsilon_{\rm cal}^2 + \epsilon_{\rm sky,local}^2 + \epsilon_{\rm sky,global}^2}
\ee
with
\be
\epsilon_{\rm sky,local} = \sigma_{\rm sky,local} \Omega_{\rm pix} \sqrt{N_{\rm pix}}
\ee
and
\be
\epsilon_{\rm sky,global} = \sigma_{\rm sky,global} \Omega_{\rm pix} N_{\rm pix}
\ee
where $\sigma_{\rm sky,local}$ is the standard deviation of the sky values in the combined set of sky apertures, $\sigma_{\rm sky,global}$ is the standard deviation of the median sky values for the set of sky apertures, $\Omega_{\rm pix}$ is the solid angle subtended per pixel, and $N_{\rm pix}$ and $N_{\rm sky}$ is the number of pixels in an annulus.  Following \cite{boselli03} and \cite{ciesla12}, $\sigma_{\rm sky,local}$ accounts for random error contributions from small-scale sky fluctuations (e.g., faint background sources) whereas $\sigma_{\rm sky,global}$ represents large-scale deviations in the sky value such as errors due to flat-fielding.  All fluxes were corrected for Galactic extinction \citep{schlafly11} assuming $A_V/E(B-V)\approx3.1$ and the reddening curve of \cite{draine03}.

\subsection{SED Fitting}
\label{sec:sed}

The full ultraviolet--infrared SEDs were fitted using the Bayesian-based CIGALE software package \citep{noll09,boquien15,burgarella15}.  This package allows the user to estimate fundamental parameters such as stellar mass, star formation rate, and the characteristic epoch of star formation and its decay rate using an energy-balanced approach whereby the diminution of ultraviolet/optical light via dust extinction is accounted for in equal amounts in the infrared via dust emission.  In our SED fitting we adopt the stellar and dust emission libraries of \cite{bruzual03} and \cite{dale14}, respectively, the \cite{chabrier03} stellar initial mass function, and a dust attenuation curve based on the work of \cite{calzetti00} and \cite{leitherer02}.  The fit parameters include metallicity, extinction, attenuation curve modifier,\footnote{The attenuation curve modifier parameter $\delta$ governs the slope of the extinction law, i.e., the baseline starburst attenuation law multiplied by a factor $\propto\lambda^\delta$, with $\delta=0$ corresponding to a starburst and $\delta<0$ yielding steeper attenuation curves such as those observed for the Magellanic Clouds \citep{noll09,boquien12}.} and characteristics of the major star-forming events.  For example, an exponential decreasing star formation history (also known as the ``$\tau$ model'') beginning at time $t_1$ with $e$-folding time $\tau_1$ and amplitude $\mathcal{A}_1$ would be expressed as
\be
SFR(t) \propto \mathcal{A}_1 e^{-(t-t_1)/\tau_1}, \;\;\;\; \mathcal{A}_1(t-t_1<0)=0
\label{eq:exponential}
\ee
\citep{papovich01,borch06,gawiser07,lee09}.  Another form of star formation history we utilize here is the so-called delayed star formation history model \citep{lee10,lee11,schaerer13}, i.e., 
\be
SFR(t) \propto \mathcal{A}_0 t e^{-(t-t_0)/\tau_0}, \;\;\; \mathcal{A}_0(t-t_0<0)=0.
\ee
Unlike the decreasing exponential star formation history for which the maximum occurs at $t-t_1=0$, in this formulation the maximum star formation rate occurs at the value of the $e$-folding rate after the onset of star formation: $t-t_0=\tau_0$.  Figure~\ref{fig:sfh} shows the characteristics of four example delayed star formation models in addition to an example of an exponential decreasing model.  Note how smaller values of $\tau_0$ correspond to more sharply defined and earlier characteristic epochs of star formation.  Both the delayed and decreasing exponential models for star formation histories are utilized in this work since they are common, simple prescriptions that rely on a small number of parameters.  CIGALE-based simulations show that the delayed star formation model provides superior accuracy in recovering galaxy stellar masses and star formation rates \citep{buat14,ciesla15}.  However, \cite{noll09} caution that the degeneracies inherent to the models result in less than well-constrained $\tau$ values for fits based on broad-band photometry of nearby galaxies; in such cases $\tau$ values can only typically be characterized as ``rather high'' or ``rather low''.

The input parameter ranges are provided in Table~\ref{tab:parameters}.  The main output physical parameters (e.g., $M_*$, $\tau$, $A_V$) are computed based on a probability distribution function (PDF) analysis: the SED fit $\chi^2$ is derived for each combination of input parameters, a PDF is constructed for each parameter based on the $\chi^2$ for the best fit models, and the probability-weighted means and standard deviations of the PDFs are adopted as the values and associated uncertainties of the output physical parameters (see \S~2.2 of \cite{noll09} for details of the process).

\section{Results}
\label{sec:results}

\subsection{Directly Observed: Fluxes}
Table~\ref{tab:fluxes} provides the integrated fluxes arising from within the $2a_{25} \times 2b_{25}$ apertures.  For the subset of the sample that has published data, spanning nine galaxies and seven wavelengths, these fluxes agree quite well with published values: the average ratio of these fluxes to literature fluxes is $0.98\pm0.02$.  Two notable exceptions are for the GALEX measurements of NGC~4625, a galaxy with known extended ultraviolet emission \citep{thilker07}.

\subsection{Directly Observed: Surface Brightness Profiles and Optical Colors}
\begin{figure}
 \plotone{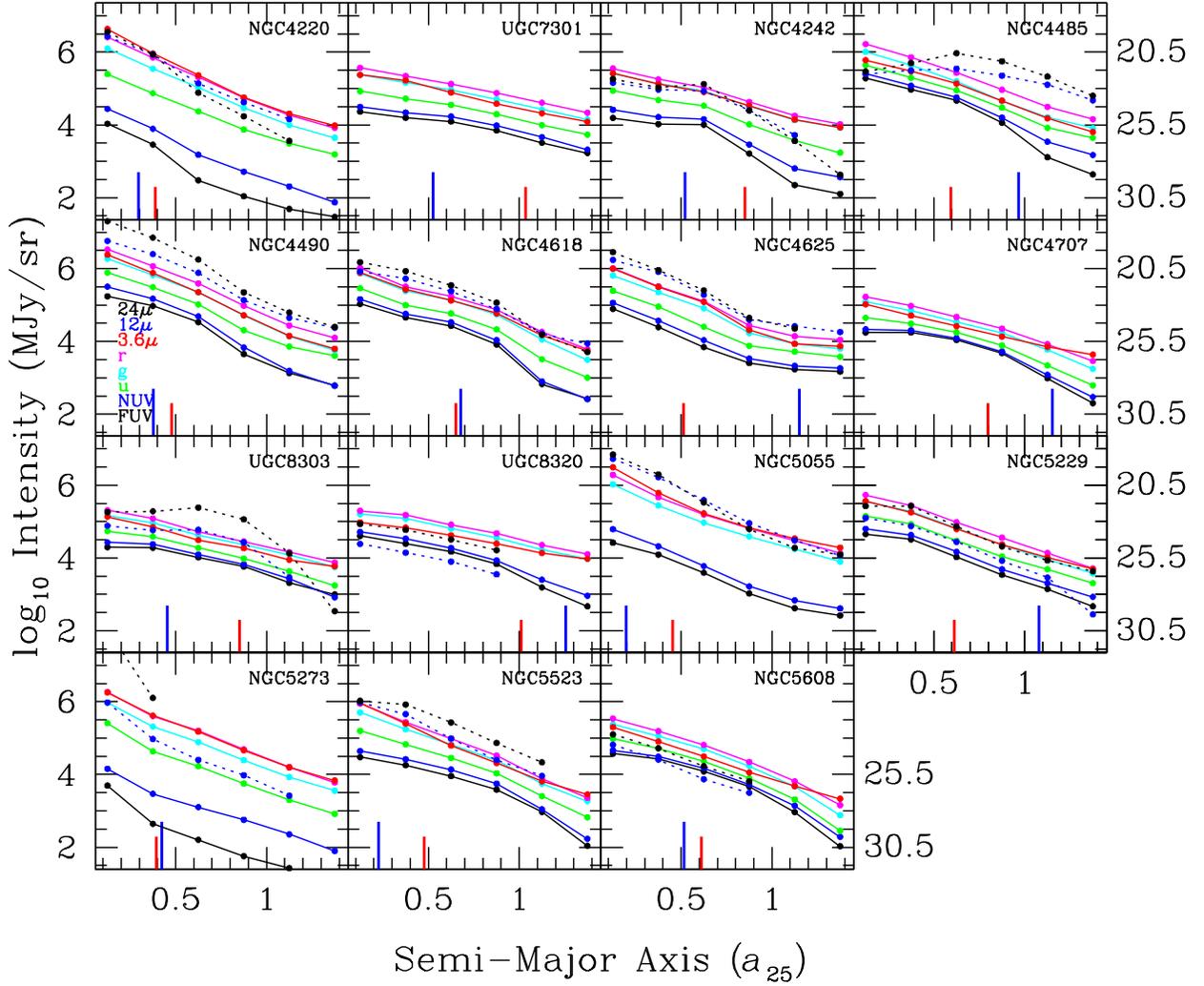}
 \caption{The multi-wavelength surface brightness profiles corrected for foreground Milky Way dust attenuation.  The right-hand axis is AB mag~arcsec$^{-2}$.  The dotted curves are for dust emission, and the vertical blue (red) lines near the bottom of each panel indicate 3~kpc (twice the $r^\prime$ half-light radius).}
 \label{fig:sb_all}
\end{figure}

Figure~\ref{fig:sb_all} displays the surface brightness profiles for the seven different wavelengths observed for each galaxy.  The profiles generally fall with radius, though for a few galaxies the peak of the warm dust emission (usually traced by MIPS 24\m) is significantly spatially displaced from the galaxy centers; the warm dust emission does not spatially mimic the stellar emission for NGC~4242, NGC~4485, UGC~8303, and NGC~5523.  The dust emission surface brightness profile is also one of two profiles that is not systematically detected sample-wide out to the last annular region of 1.25$<$$a/a_{25}$$<$1.50; the (detection of the) dust emission is truncated for NGC~4420, NGC~4242, NGC~4625, UGC~8320, NGC~5273, and NGC~5608 (the far-ultraviolet is also undetected for the last annular region of NGC~4220).  Many galaxies show fairly consistent profile shapes in the ultraviolet/optical/near-infrared, but we note that the two most inclined galaxies in our sample, UGC~7301 ($b/a=0.13$) and NGC~5229 ($b/a=0.17$), exhibit the smallest scatters in the surface brightness profile (log-linear) slopes $m(\lambda) = \Delta \log I_\nu(\lambda) / \Delta (a/a_{25})$, with $\sigma(m) = 0.10$~dex and 0.08~dex for UGC~7301 and NGC~5229, respectively.  This similarity in slopes across wavelengths for the more inclined systems is not the result of reddening effects, since the internal extinctions derived in the SED fitting (\S~\ref{sec:sed}) are not unusually large for UGC~7301 and NGC~5229.  Rather, the similarity is more likely due to the projected mixing of inner and outer disk regions for the central annular apertures.  The less inclined galaxies show nuanced differences in their multi-wavelength surface brightness profiles, which in turn lead to variations in the color profiles and in the interpreted star formation histories, as discussed below.

\cite{pohlen06} studied the $g^\prime$ and $r^\prime$ surface brightness profiles of 85 moderately inclined, late-type spiral galaxies from SDSS.  \cite{martinnavarro12} carried out a similar study for 34 inclined galaxies using SDSS and {\it Spitzer} 3.6\m\ images.  \cite{pohlen06} found that 60\% (30\%) of their systems displayed exponential disks followed by downbending (upbending) surface brightness profiles; \cite{martinnavarro12} found generally similar results.  They saw downbending features primarily for later-type spiral galaxies and upbending for earlier-type spirals. Our relatively coarse annular sampling may not yield surface brightness profiles as detailed as those studied by \cite{pohlen06}, but we can explore any trends near the galaxy outskirts with higher signal-to-noise, especially since our images go significantly deeper (\S~\ref{sec:new}).  Our multi-wavelength collection of surface brightness profiles in Figure~\ref{fig:sb_all} exhibit all three types of profiles.  However, only two of our targets satisfy the \cite{pohlen06} joint criteria of late-type spiral morphology ($3\leq T \leq 8$) and moderately inclined ($b/a>0.5$), so no strong conclusions based on going significantly deeper than SDSS can be drawn here.

\begin{figure}
 \plotone{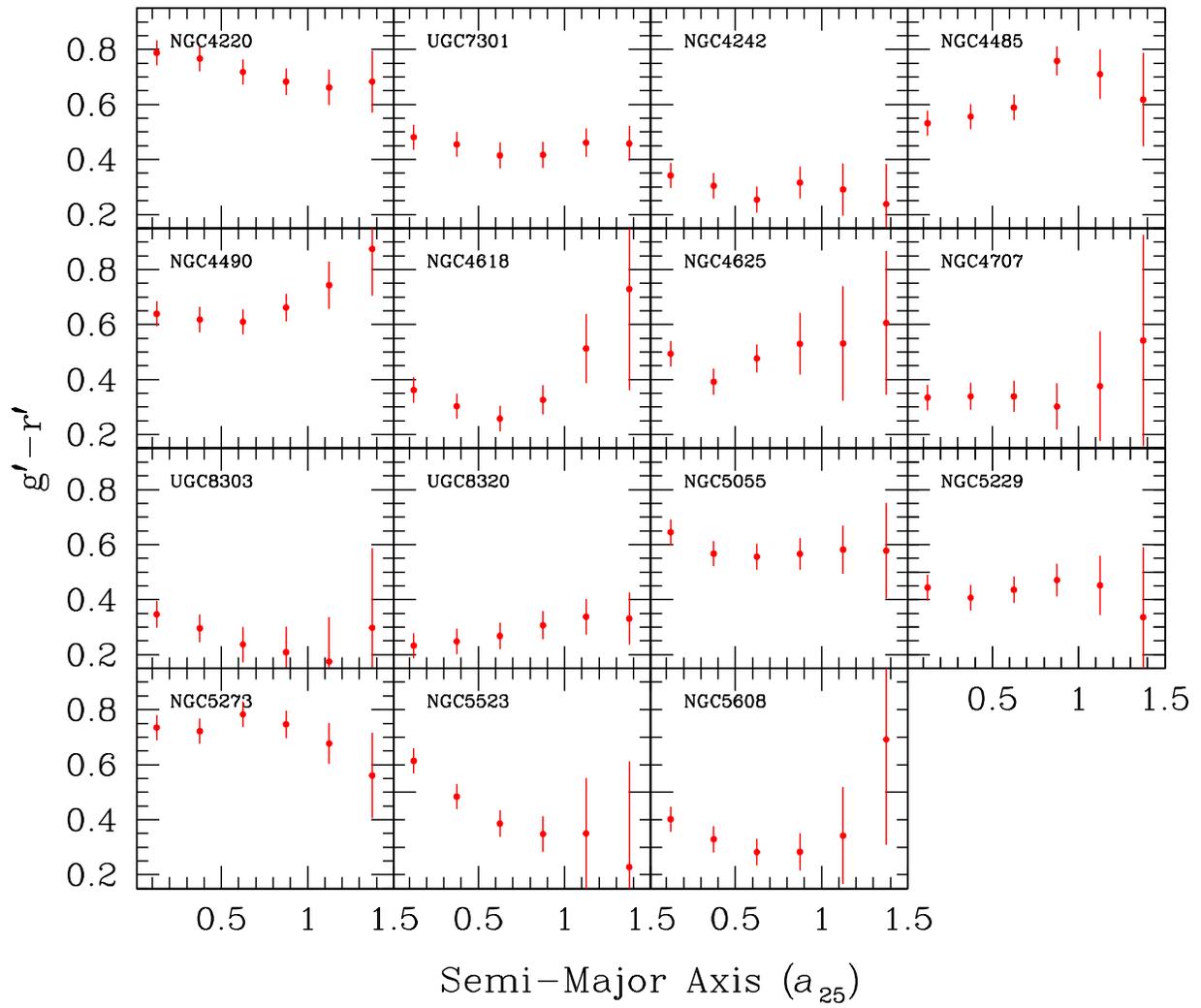}
 \caption{The $g^\prime - r^\prime$ color radial profiles.}
 \label{fig:gr_all}
\end{figure}

On the other hand, stacking the images of many galaxies can yield exquisitely sensitive surface brightness profiles \citep[e.g.,][]{zibetti04,tal11}.  Though such analyses cannot provide insight on individual systems, they do inform us about the broad brush characteristics for large ensembles of galaxies.  \cite{dsouza14} stack the $g^\prime$ and $r^\prime$ SDSS images for over 45,000 isolated galaxies at redshifts $0.05 \leq z \leq 0.1$.  Binned according to global galaxy stellar mass, their stacks probe down to an effective surface brightness of $\mu(r) \sim 32$~mag~arcsec$^{-2}$ \citep[compared to 27~mag~arcsec$^{-2}$ for][]{pohlen06}.  They find that the outer portions, which they refer to as the stellar halos, are redder than the main disk for all systems with $M_*<10^{11}~L_\odot$, and that the detected halo mass fraction rises with increasing galaxy mass.  This reddening feature is not due to effects of dust attenuation since galaxy dust mass surface densities are minimal for larger radii \citep{munozmateos09a,aniano12}.  Figure~\ref{fig:gr_all} presents a compilation of the radial $g^\prime - r^\prime$ trends for our sample.  Our galaxies exhibit a similar range in $g^\prime - r^\prime$ color as seen in the SDSS stacks \citep[see also][]{west09,tortora10}.  Considered as a whole, our sample shows an outward reddening color trend (upper panel of Figure~\ref{fig:tau_gr}); on average, our galaxies are redder at $a=1.38a_{25}$ by 0.08~mag in $g^\prime - r^\prime$ compared to the value at $a=0.63a_{25}$.  \cite{dsouza14} suggest such peripheral reddening implies accretion of old stars into the stellar halo.  However, given the typical color uncertainties in the outskirts, this average reddening is modest and is clearly seen for only a few individual galaxies (e.g., NGC~4490, NGC~4618, NGC~4707, NGC~5608).  There is a considerable diversity in the galaxy color profiles, with some galaxies exhibiting flat (NGC~5055, UGC~7301) or even blue trends (NGC~5273, NGC~5523), and so caution must be taken when analyzing sample-wide averages.

\begin{figure}
 \plotone{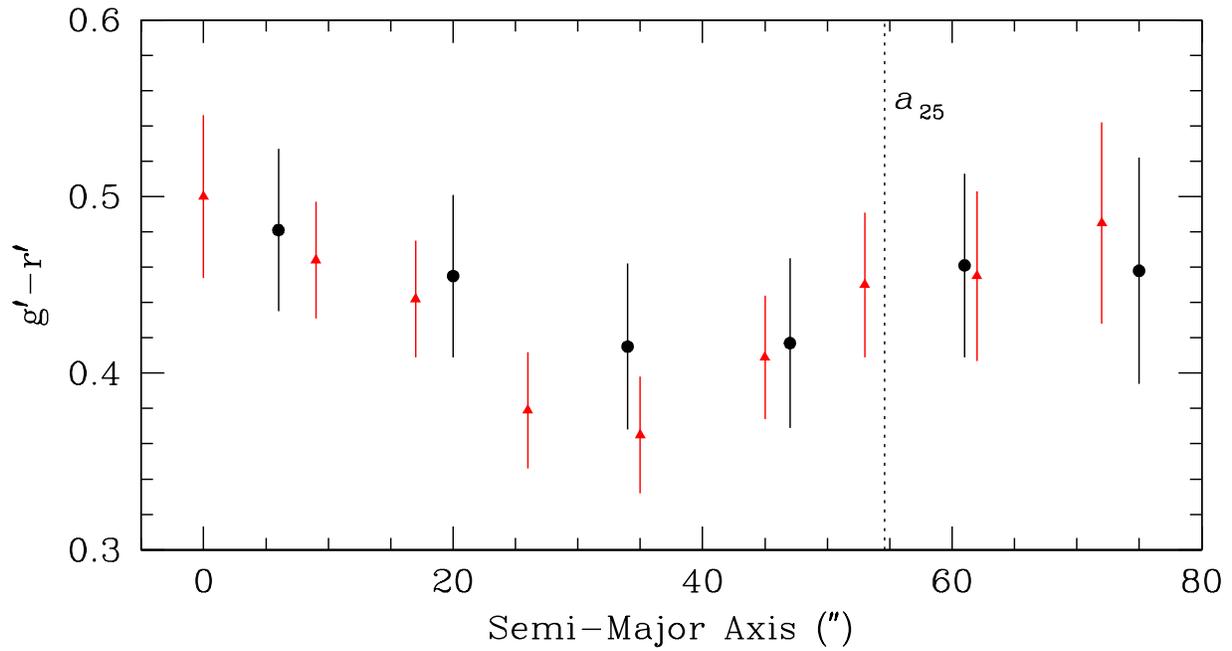}
 \caption{Comparison of annular-based optical colors (circles) with those derived from a series of 9\arcsec-diameter apertures placed along the major axis of UGC~7301 (triangles).}
 \label{fig:u7301}
\end{figure}

It has been shown that the extended wings of the Point Spread Function (PSF) can lead to artificially red optical colors for observations utilizing thinned CCDs \citep{michard02}.  This so-called ``red halo'' effect stems from increased instrumental scattering at longer wavelengths for radial distances greater than $\sim$15\arcsec, leading some authors to ignore SDSS $i$ band data when studying galaxy colors: the effect is much less pronounced for the shorter-wavelength SDSS filters while the SDSS $z$ band data are taken with unthinned CCDs and thus do not suffer from this effect \citep{wu05,liu09}.  Though the WIROPrime camera utilizes a thinned detector, $i$ band data were not utilized in this analysis.  A second potential PSF-related issue is that each line-of-sight contains emission from a range of galacto-centric distances for inclined galaxies, and the impact is of course largest for edge-on galaxies.  We investigate the potential impact of both inclination and the red halo effect for UGC~7301, the smallest and most inclined galaxy in the sample.  Figure~\ref{fig:u7301} provides a comparison of $g^\prime - r^\prime$ colors for our annuli and for a series of 9\arcsec-diameter apertures placed along the major axis, an arrangement that promotes a more PSF-independent radial analysis.  No obvious differences appear between the two color trends, and so annular smearing is expected to be negligible in this work.  To enable further interpretation of the observed colors, we turn to SED fitting of the complete collection of panchromatic surface brightness profiles.

\subsection{Inferred from SED Fitting: Star Formation Histories}
\label{sec:sfh}

\begin{figure} 
 \plotone{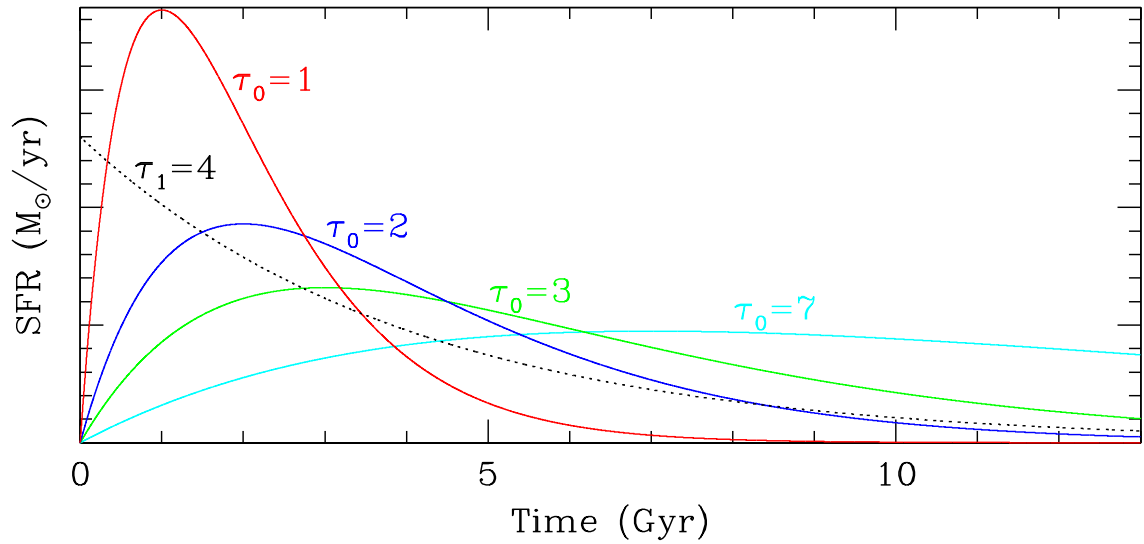}
 \caption{Four examples of a ``delayed'' star formation history along with an example of a decreasing exponential star formation history (dotted curve).  All values shown are in Gyr.}
 \label{fig:sfh}
\end{figure}

\begin{figure}
 \plotone{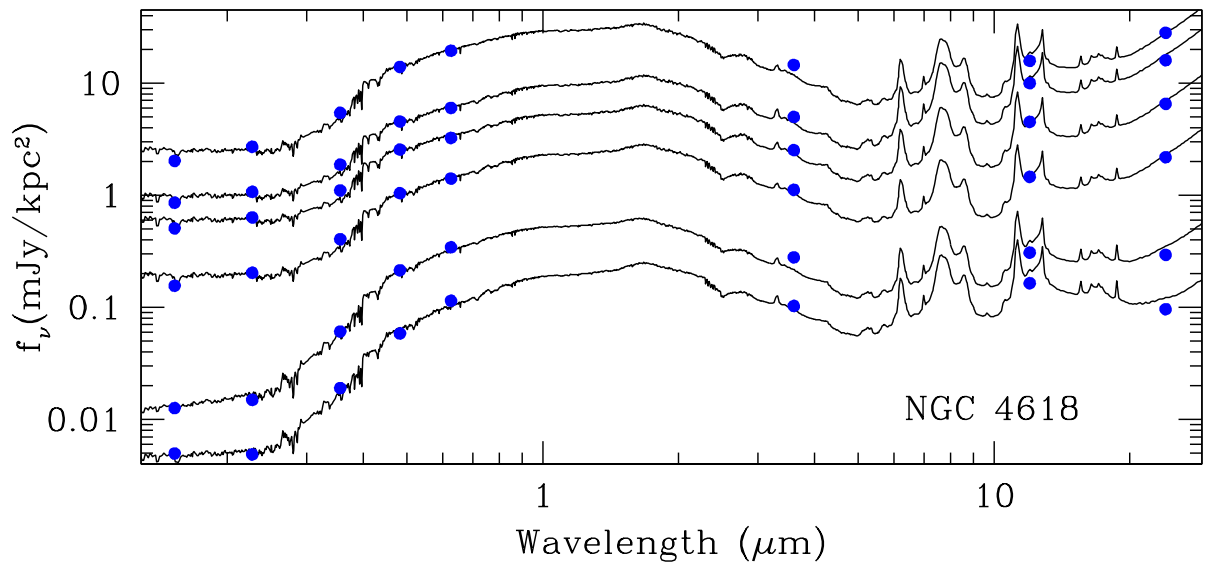}
 \caption{SEDs for the six annular regions of NGC~4618.  The blue dots indicate the measured surface brightnesses and the black curves show the best-matched stellar+dust SEDs assuming a delayed star formation history.}
 \label{fig:sed}
\end{figure}

\begin{figure}
 \plotone{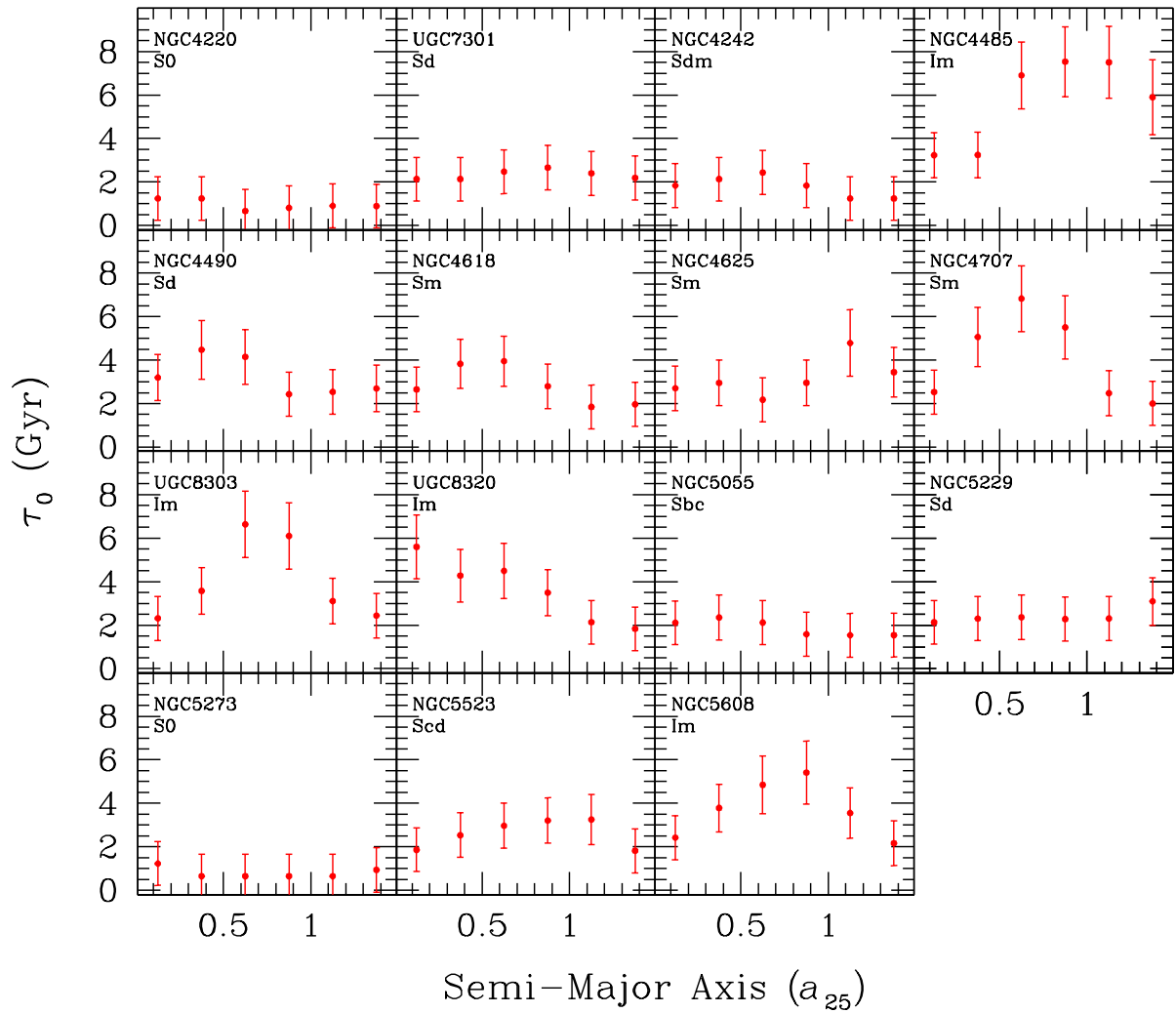}
 \caption{The $e$-folding $\tau_0$ values assuming a delayed star formation history.}
 \label{fig:tau_all}
\end{figure}

An example of our SED fitting results is displayed in Figure~\ref{fig:sed} for NGC~4618, a galaxy for which dust emission is detected throughout the six annular regions studied.  As described in \S~\ref{sec:sed}, multiple parameters are involved in such fits.  Figure~\ref{fig:tau_all} presents for each galaxy the $e$-folding timescale radial trends for a delayed star formation history.  We experimented with allowing $t_0$, the age of the oldest stars, to be a free parameter with a range of 6--13~Gyr ago.  Since the radial trends for $\tau_0$ were essentially unchanged for different values of $t_0$, we adopted a fixed value of $t_0=11$~Gyr ago in order to minimize the number of free parameters in the fits.  As a result of fixing $t_0$, any differences in $\tau_0$ directly correspond to differences in the timing of the peak of the star formation histories (see Figure~\ref{fig:sfh} and the corresponding discussion in \S~\ref{sec:sed}).  The error bars seen in Figure~\ref{fig:tau_all} are generally larger for larger values of $\tau_0$, which can be visually deciphered from inspection of Figure~\ref{fig:sfh}: the profile of the delayed star formation model is broader for larger $\tau_0$ and thus naturally leads to a less well constrained epoch for the peak in the star formation history.  

\begin{figure}
 \plotone{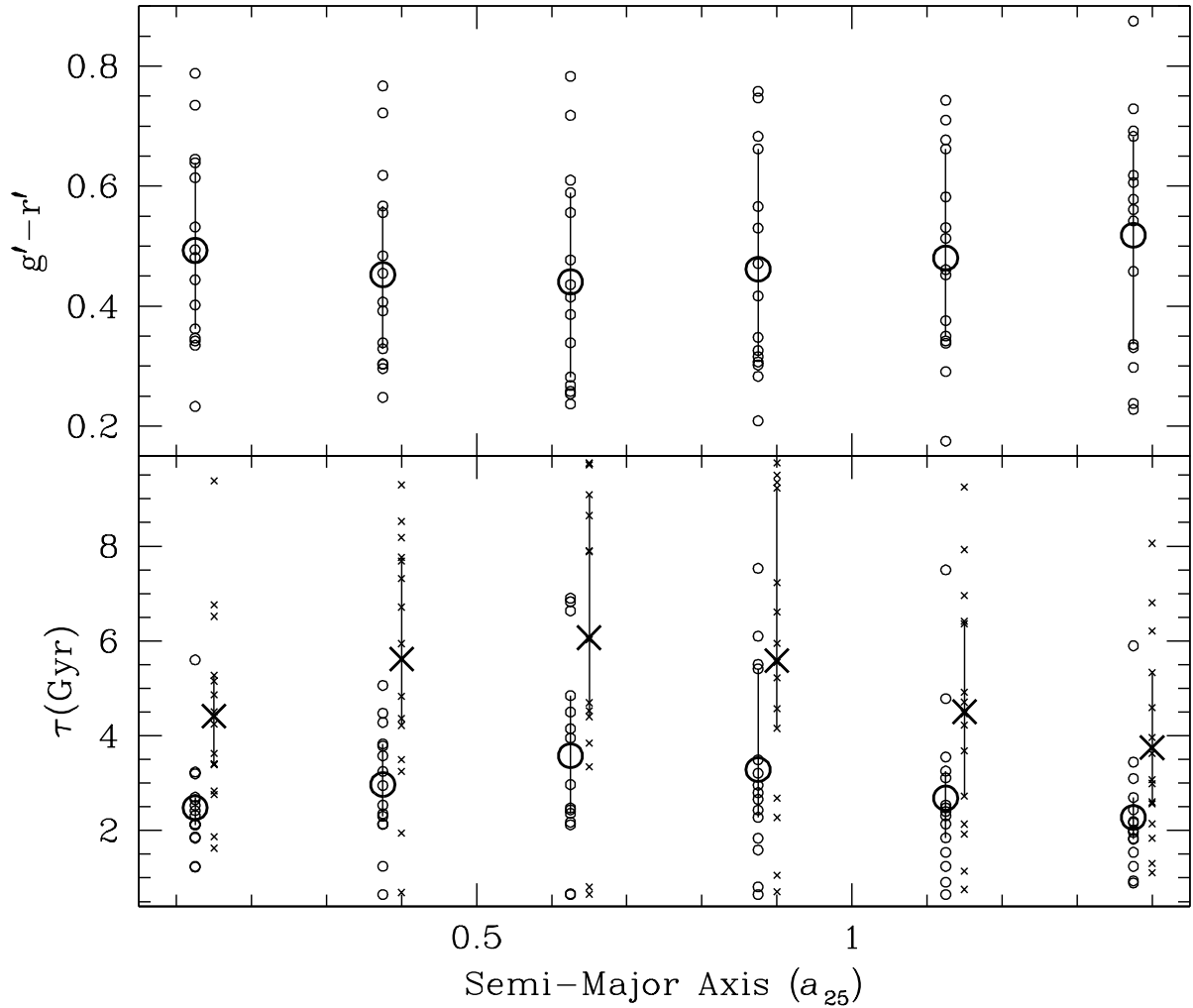}
 \caption{The distribution of $g^\prime - r^\prime$ and $\tau$ parameters.  The large symbols indicate the average value for each radius; in the bottom panel the circles stem from a delayed star formation history and the (slightly offset for clarity) crosses indicate the single exponential star formation history model.  The vertical lines mark the 25$^{\rm th}$--75$^{\rm th}$ quartile spreads.}
 \label{fig:tau_gr}
\end{figure}

When all the $\tau_0$ values are presented in a single plot (lower panel of Figure~\ref{fig:tau_gr}), the sample overall shows older stellar populations stemming from more sharply defined epochs of star formation at both the galaxy centers and galaxy outskirts compared to the stellar populations at mid-galacto-centric distances.  On average, our galaxies' characteristic epoch of star formation is $\sim 1-2$~Gyr earlier at $r=0.13a_{25}$ and $1.38a_{25}$ compared to the value at $r=0.63a_{25}$.  Similar results are seen for the $e$-folding time $\tau_1$ when a decreasing exponential decreasing star formation history is considered (crosses in Figure~\ref{fig:tau_gr}).  Nonetheless, there is a large diversity in the individual profiles, similar to what is seen for the $g^\prime - r^\prime$ colors; for some galaxies the $\tau_0$ trend is essentially flat, for others it is mildly falling, there are galaxies that show rising-then-falling radial trends, etc.  Thus, it is perhaps more illuminating to note the sample diversity than any average trend that may be unduly influenced by outliers, especially in light of the range of morphologies represented in the sample.  The two S0 galaxies, NGC~4220 and NGC~5273, are notable in this presentation since they show the smallest and most constant values for $\tau_0$.  The implication is that the bulk of their stars formed spatially more uniformly across the disk, and with an earlier and more sharply defined epoch of star formation, than the stellar populations in the 13 other galaxies in the sample.  

About half of the sample exhibits trends suggesting older on both the inside and the outside, consistent with being redder in those locations (provided our constraints on dust reddening and metallicity are reasonable).  This scenario echoes that seen observationally for M~33 and other nearby galaxies: an inside-out formation process for the main galaxy disk where the stars are younger and the colors increasingly blue with radius \citep[e.g.,][]{gonzalezdelgado14}, followed by a redder and older stellar population in the galaxy peripheries or haloes \citep{williams09,gogarten10,barker11}.  \cite{bullock05} find in their simulations of galaxy stellar halo formation that the bulk of the mass in haloes derives from the remnants of accreted satellites merger tidal debris that occurred on average 9~Gyr ago.  During the time that has passed since those accretion events, the stars have passively evolved to provide an old, red present-day appearance for the spiral galaxy haloes.  Alternatively, \cite{roskar08} and \cite{sanchezblazquez09} predict that radial migration, a process whereby the influence of passing spiral arms can transport stars outwards, results in a conspicuous old--young--old mean radial age profile, in essence creating a red stellar periphery through internal dynamical processes.  Though there is ample evidence that the outer disks of many galaxies possess pockets of active star formation \citep{gildepaz05,thilker07,dong08,alberts11}, there are not enough OB stars in galaxy peripheries to result in overall blue stellar haloes.   

\begin{figure}
 \plotone{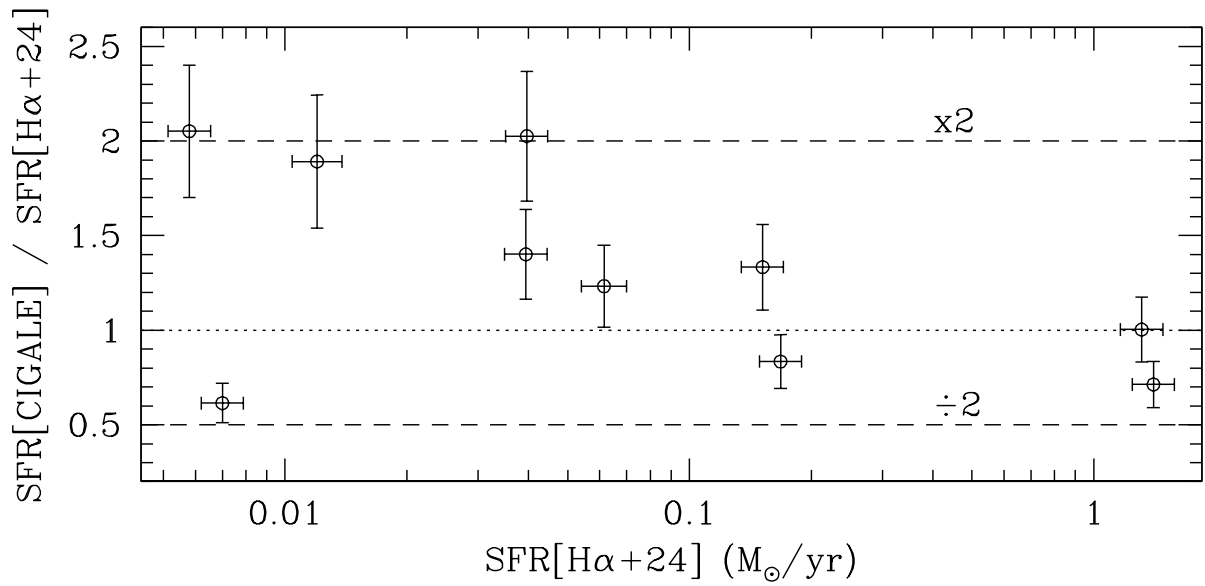}
 \caption{Comparison between the global star formation rates output by our CIGALE fits and those from a prescription that utilizes global H$\alpha$ and 24\m\ data.}
 \label{fig:sfr}
\end{figure}

\begin{figure}
 \plotone{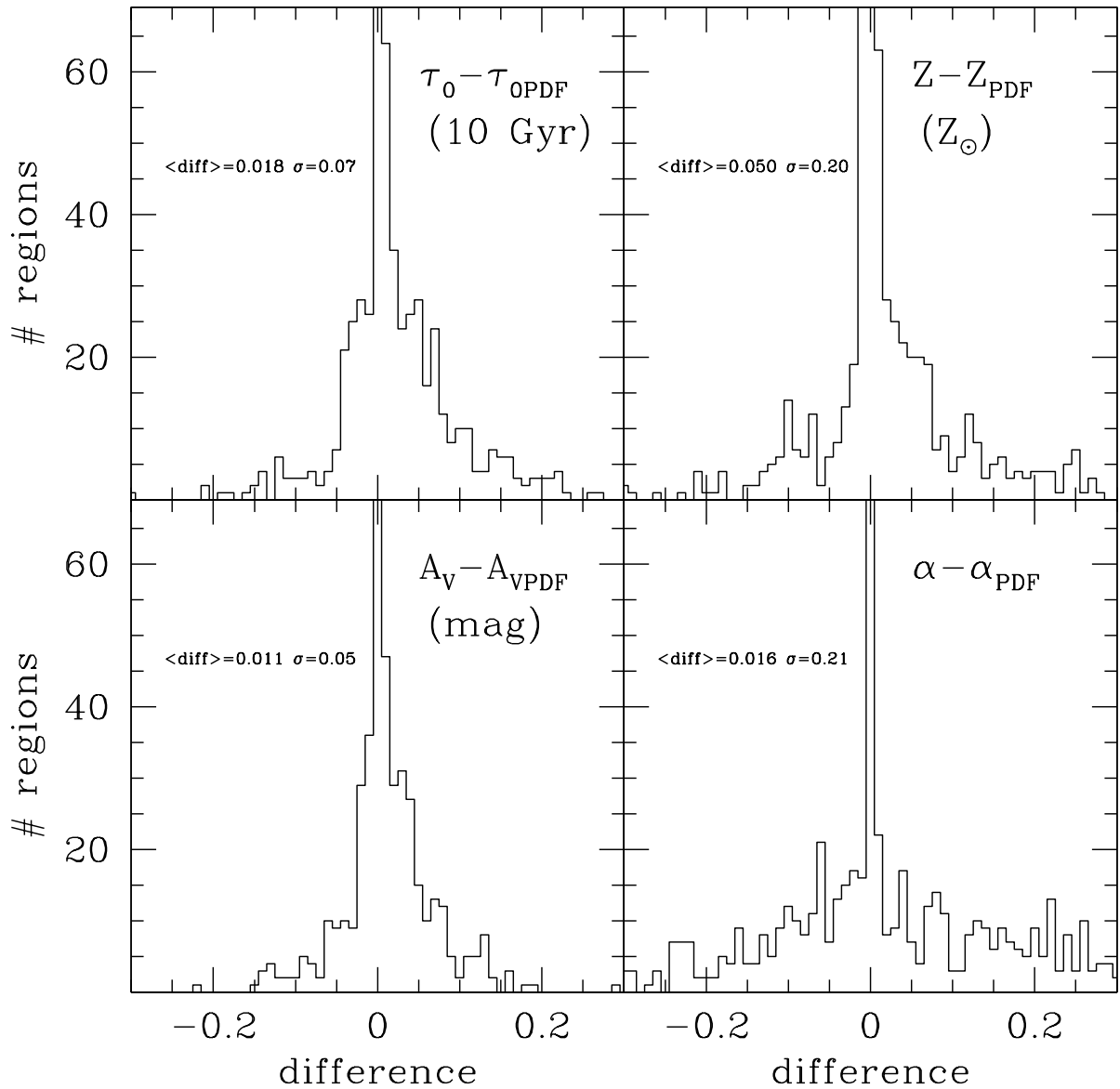}
 \caption{Comparison between Bayesian-based best-fit parameters with the fit results for a series of Monte Carlo simulations that inject uncertainty into the measured fluxes.}
 \label{fig:mc}
\end{figure}

While we are mainly interested in the $e$-folding timescales inferred from the SED fitting, it is important to verify that the other output physical parameters derived are reasonable, as a check on the overall fit reliability.  The fitted metallicities range from half-solar to slightly super-solar, with higher values preferentially found near the galaxy optical centers.  The extinction range is 0.1--1.1~mag in $A_V \approx 3.1 E(B-V)_*^y$ and again the average value peaks, as expected, for the innermost annular regions.  Figure~\ref{fig:sfr} shows a comparison between the global star formation rates output from the CIGALE fits and those independently derived from a combination of the global H$\alpha$ and 24\m\ data, taking care to ensure that both estimates use the same galaxy distances and initial mass function prescriptions \citep[e.g., Equations~1 and 16 of][]{kennicutt08}.  The average ratio of the two types of star formation rates is 0.95, with a dispersion of 0.4 and peak-to-peak variations within a factor of $\sim$2 from unity.  Detailed studies have also been carried out that compare simulated input parameters for a sample of mock galaxies with the output parameters extracted from CIGALE SED fitting. \citep{giovannoli11,boquien12,ciesla15}.  \cite{ciesla15}, for example, find that CIGALE returns stellar masses and star formation rates that are systematically low by 5--10\% (for decreasing exponential and delayed star formation histories), while \cite{giovannoli11} and \cite{boquien12} respectively find that the shape of the dust SED and the slope of the power law that modifies the attenuation curve ($\alpha$ and $\delta$ in Table~\ref{tab:parameters}) are the least accurate parameters returned.  To help further assess the validity of the SED fit parameters, we have carried out a series of Monte Carlo simulations.  In each simulation a random flux offset was added to each flux, with the flux offset for a given wavelength and radial position derived from a Gaussian distribution with $\sigma$ scaled according to the measured uncertainty at that same wavelength and radial position.  For each of these simulations the same SED fitting and Bayesian-based analysis described above was carried out.  Figure~\ref{fig:mc} presents a comparison between the standard fit parameters and those from the Monte Carlo analysis.  The standard deviation of the difference between the canonical and simulated $e$-folding time is about 700~Myr, a reasonable value given the estimated uncertainties.  The output parameters with the broadest distributions in Figure~\ref{fig:mc} are the dust model power-law parameter $\alpha$ \citep[see][]{dale14} and the stellar metallicity $Z$: the simulated values for these parameters disperse noticeably more from the standard values, with standard deviations in each distribution of $\sim$0.2.

\section{Summary and Discussion}
\label{sec:summary}

The {\it Spitzer} EDGES survey provided large and extremely sensitive near-infrared maps for 92 nearby galaxies.  We report here results for a subset of the EDGES sample based on EDGES and ancillary data in addition to a follow-up ground-based campaign for deep $u^\prime$$g^\prime$$r^\prime$ imaging on the 2.3~m WIRO telescope.  This panchromatic database is utilized to study radial trends in galaxy surface brightnesses, colors, and star formation histories, with a primary goal of taking advantage of the imaging depth in the outer portions ($<1.5 a_{25}$) of the galaxies.  The star formation histories are estimated using ultraviolet-optical-infrared SED fits executed in an energy-balanced fashion, whereby the ultraviolet/optical radiation attenuated by dust is converted in equal portions to dust emission in the infrared.  While most surface brightness profiles peak at the galaxy centers and then systematically fall with radius, the dust profiles for four late-type systems peak off-center.  In these systems the $g^\prime - r^\prime$ radial color trends are generally bluer, and the star formation history profiles younger, where the warm dust emission peaks suggesting that the dust is tracing sites of recent star formation.  For a subset of the sample we find results similar to those from SDSS-based analyses of optical galaxy colors \citep{dsouza14}, where both the central and the outer galaxy regions show redder colors than the mid-galactic radial regions.  The star formation histories provide additional evidence, suggesting for this subset that these mid-galactic regions are on average younger than the central bulge and the galaxy peripheries.  These results are consistent with disks forming in an inside-out fashion combined with red stellar outskirts formed through contributions from either radial migration or the cumulative effect of past mergers and accretion events.  However, there are significant variations on a galaxy-by-galaxy basis and thus a larger study encompassing a much larger fraction of the EDGES sample is warranted.

\acknowledgements 
We thank the referee for excellent suggestions.
DAD thanks IPAC/Caltech and the Laboratoire d'Astrophysique de Marseille for their hospitality during the course of this work.
This work is supported by the National Science Foundation under REU grant AST 1063146 and by NASA through an award issued by JPL/Caltech.
This work is based on observations made with the {\it Spitzer Space Telescope} and utilizes the NASA/IPAC Infrared Science Archive, both operated by JPL/Caltech under a contract with NASA.  {\em Herschel} is an ESA space observatory with science instruments provided by European-led Principal Investigator consortia and with important participation from NASA. We gratefully acknowledge NASA's support for construction, operation, and science analysis for the GALEX mission, developed in cooperation with the Centre National d'Etudes Spatiales of France and the Korean Ministry of Science and Technology.
Funding for the Sloan Digital Sky Survey and SDSS-II has been provided by the Alfred P. Sloan Foundation, the Participating Institutions, the NSF, the U.S. Department of Energy, NASA, the Japanese Monbukagakusho, the Max Planck Society, and the Higher Education Funding Council for England.

\bibliography {dad}
\begin{deluxetable}{llclrrcr}
\tablenum{1}

\def\p{$\pm$}
\tabletypesize{\scriptsize}
\tablecaption{Galaxy Sample}
\tablewidth{0pc}
\tablehead{
\colhead{Galaxy} &
\colhead{Alternative} &
\colhead{$\alpha_0$~\&~$\delta_0$} &
\colhead{Optical} &
\colhead{$2a_{25} \times 2b_{25}$} &
\colhead{c$z$} &
\colhead{$A_V$} &
\colhead{P.A.}
\\
\colhead{} &
\colhead{Name} &
\colhead{(J2000)} &
\colhead{Morphology} &
\colhead{(~$^\prime~\times~^\prime$~)} &
\colhead{(km~s$^{-1}$)} &
\colhead{(mag)} &
\colhead{($\degr$)}
}
\startdata
NGC4220 &UGC7290     &121611.7$+$475300&SA0         & 3.89$\times$1.36& 914&0.049&139.7\\
UGC7301 &            &121642.1$+$460444&Sd          & 1.82$\times$0.24& 690&0.030& 81.6\\
NGC4242 &UGC7323     &121730.2$+$453709&SAB(s)dm    & 5.01$\times$3.81& 506&0.033& 26.9\\
NGC4485 &UGC7648     &123031.1$+$414204&IB(s)m pec  & 2.29$\times$1.63& 493&0.059&  3.8\\
NGC4490 &UGC7651     &123036.2$+$413838&SB(s)d pec  & 6.31$\times$3.09& 565&0.060&121.2\\
NGC4618 &UGC7853     &124132.8$+$410903&SB(rs)m     & 4.17$\times$3.38& 544&0.058& 26.9\\
NGC4625 &UGC7861     &124152.7$+$411626&SAB(rs)m pec& 2.19$\times$1.90& 621&0.050&133.4\\
NGC4707 &DDO150      &124822.9$+$510953&Sm?         & 2.24$\times$2.08& 468&0.030& 23.1\\
UGC8303 &HolmbergVIII&131317.6$+$361303&IAB(s)m     & 2.24$\times$1.90& 944&0.049&177.1\\
UGC8320 &DDO168      &131427.9$+$455509&IBm         & 3.63$\times$1.38& 192&0.042&149.9\\
NGC5055 &Messier63   &131549.3$+$420145&SA(rs)bc    &12.59$\times$7.18& 484&0.048& 94.0\\
NGC5229 &UGC8550     &133402.8$+$475456&SB(s)d?     & 3.31$\times$0.56& 364&0.049&166.8\\
NGC5273 &UGC8675     &134208.3$+$353915&SA0(s)      & 2.76$\times$2.51&1085&0.028&  3.5\\
NGC5523 &UGC9119     &141452.3$+$251903&SA(s)cd?    & 4.57$\times$1.28&1039&0.052& 92.0\\
NGC5608 &UGC9219     &142317.9$+$414633&Im?         & 2.63$\times$1.34& 663&0.026& 94.9\\

\enddata
\tablecomments{\footnotesize The apertures used for the photometry have the centers and position angles (measured east of north) listed here, with ellipticities determined via $b_{25}/a_{25}$, where $2a_{25}$ and $2b_{25}$ are respectively the RC3 major axis and minor axis sizes of the $B$ band isophote defined at 25~mag~arcsec$^{-2}$.  All information is taken from the NASA/IPAC Extragalactic Database (NED) including the foreground Milky Extinction.}
\label{tab:sample}
\end{deluxetable}

\begin{deluxetable}{lrrcccccr}
\tablenum{2}
\def\a{\tablenotemark{a}}
\def\b{\tablenotemark{b}}

\def\p{$\pm$}
\tabletypesize{\scriptsize}
\tablecaption{Imaging Integrations}
\tablewidth{0pc}
\tablehead{
\colhead{Galaxy} &
\colhead{\it GALEX} &
\colhead{\it GALEX} &
\colhead{WIRO} &
\colhead{WIRO} &
\colhead{WIRO} &
\colhead{\it Spitzer} &
\colhead{WISE} &
\colhead{\it Spitzer} 
\\
\colhead{} &
\colhead{FUV} &
\colhead{NUV} &
\colhead{$u^\prime$} &
\colhead{$g^\prime$} &
\colhead{$r^\prime$} &
\colhead{3.6\m} &
\colhead{12\m} &
\colhead{24\m} 
}
\startdata
NGC4220 &  12101&12101&   3600&3600&3600&1800& 72& 10\\
UGC7301 &  13756&13757&   3600&3600&3600&1800& 72& 72\a \\
NGC4242 &   1683& 3282&   3600&3600&3600&1800& 72&160\\
NGC4485 &   1587& 2810&   3600&3600&3600&1800& 72&160\\
NGC4490 &   1587& 2810&   3600&3600&3600&1800& 72&160\\
NGC4618 &   3242& 3259&   3600&3600&3600&1800& 72&160\\
NGC4625 &   3259& 3259&   3600&3600&3600&1800& 72&160\\
NGC4707 &   1664& 1664&   3600&3600&3600&1800& 72&160\\
UGC8303 &    205& 1623&   3600&3600&3600&1800& 72& 72\a \\  
UGC8320 &   1576& 3116&\nodata&3600&3600&1800& 72&160\\
NGC5055 &   3002& 3754&\nodata&3600&3600&1800& 72&160\\ 
NGC5229 &   2757& 2757&   3600&3600&3600&1800& 72&160\\
NGC5273 &   1659& 1659&   3600&3600&3600&1800& 72&282\b \\ 
NGC5523 &     96&   96&   3600&3600&3600&1800& 72& 72\a \\
NGC5608 &    105&  105&   3600&3600&3600&1800& 72& 72\a \\

\enddata
\tablecomments{Integrations are in seconds per position on the sky.  The post-processing images utilized here have resolutions of $\approx$ 6\arcsec\ (GALEX, WISE 12\m, Spitzer 24\m, Herschel 70\m), 1\farcs7 (Spitzer 3.6\m), and 2\farcs0 (WIRO).}
\tablenotetext{a} {{\it WISE} 22\m}
\tablenotetext{b} {{\it Herschel} 70\m}
\label{tab:integrations}
\end{deluxetable}

\begin{deluxetable}{lll}
\tablenum{3}
\tabletypesize{\scriptsize}
\tablecaption{Fit Parameters}
\tablewidth{0pc}
\tablehead{
\colhead{Parameter} &
\colhead{Notation} &
\colhead{Allowed Values} 
}
\startdata
Metallicity	& $Z$	& 0.008, 0.02, 0.05 \\
IMF		&	& Chabrier \\ 
Color excess: young stars &$E(B-V)_*^{\rm y}$ & 0.0, 0.025, 0.05, 0.1, 0.15, 0.2, 0.25, 0.3, 0.4	 \\
Color excess: old stars &$E(B-V)_*^{\rm o}$ & 0.44$E(B-V)_*^{\rm y}$	 \\
Dust emission template & $\alpha$	& 0.5, 1.0, 1.25, 1.50, 1.75, 2.0, 2.25, 2.50, 3.00\\
Slope of power law that modifies attenuation curve&$\delta$ & $-$0.5, $-$0.4, $-$0.3, $-$0.2, $-$0.1, 0 \\
\hline
\multicolumn{3}{c}{Delayed Star Formation History} \\
\hline
SFR $e$-folding time (Gyr)	& $\tau_0$ & 0.5, 1, 1.5, 2, 2.5, 3, 3.5, 4, 4.5, 5, 5.5, 6, 6.5, 7, 7.5, 8, 10\\
Age of oldest stars (Gyr ago) & $t_0$ & 11\\	
\hline
\multicolumn{3}{c}{Single Exponential Decreasing Star Formation History} \\
\hline
SFR $e$-folding time (Gyr)	      & $\tau_1$ & 0.5, 1, 1.5, 2, 2.5, 3, 3.5, 4, 4.5, 5, 5.5, 6, 6.5, 7, 7.5, 8, 10\\
Age of oldest stars (Gyr ago)   & $t_1$ & 11\\	

\enddata
\label{tab:parameters}
\end{deluxetable}

\begin{deluxetable}{lrrcccccr}
\tablenum{4}
\def\a{\tablenotemark{a}}
\def\b{\tablenotemark{b}}
\def\c{\tablenotemark{c}}

\def\p{$\pm$}
\tabletypesize{\scriptsize}
\tablecaption{Integrated Fluxes}
\tablewidth{0pc}
\tablehead{
\colhead{Galaxy} &
\colhead{\it GALEX} &
\colhead{\it GALEX} &
\colhead{WIRO} &
\colhead{WIRO} &
\colhead{WIRO} &
\colhead{\it Spitzer} &
\colhead{WISE} &
\colhead{\it Spitzer} 
\\
\colhead{} &
\colhead{FUV} &
\colhead{NUV} &
\colhead{$u^\prime$} &
\colhead{$g^\prime$} &
\colhead{$r^\prime$} &
\colhead{3.6\m} &
\colhead{12\m} &
\colhead{24\m} 
}
\startdata
NGC4220  &494\p050E$-$3&142\p014E$-$2&154\p008E$-$1&712\p035E$-$1&143\p007E$+$0&196\p009E$+$0&138\p014E$+$0&147\p017E$+$0\\
UGC7301  &415\p041E$-$3&567\p056E$-$3&129\p007E$-$2&344\p017E$-$2&518\p026E$-$2&311\p015E$-$2&\nodata      &\nodata      \\
NGC4242  &891\p089E$-$2&140\p014E$-$1&422\p051E$-$1&123\p006E$+$0&159\p008E$+$0&120\p006E$+$0&847\p106E$-$1&109\p017E$+$0\\
NGC4485  &124\p012E$-$1&163\p016E$-$1&284\p014E$-$1&571\p028E$-$1&979\p049E$-$1&419\p021E$-$1&961\p097E$-$1&201\p015E$+$0\\
NGC4490  &550\p056E$-$1&875\p087E$-$1&203\p010E$+$0&449\p022E$+$0&807\p040E$+$0&519\p026E$+$0&152\p015E$+$1&428\p030E$+$1\\
NGC4618  &261\p026E$-$1&333\p033E$-$1&632\p038E$-$1&158\p008E$+$0&213\p011E$+$0&167\p008E$+$0&258\p026E$+$0&397\p030E$+$0\\
NGC4625\c&409\p043E$-$2&594\p059E$-$2&135\p013E$-$1&344\p017E$-$1&535\p030E$-$1&493\p024E$-$1&101\p010E$+$0&127\p009E$+$0\\
NGC4707  &329\p033E$-$2&374\p039E$-$2&599\p040E$-$2&136\p009E$-$1&187\p011E$-$1&115\p006E$-$1&\nodata      &\nodata      \\
UGC8303  &351\p035E$-$2&423\p042E$-$2&691\p062E$-$2&174\p013E$-$1&219\p011E$-$1&137\p007E$-$1&134\p024E$-$1&511\p075E$-$1\a\\
UGC8320  &533\p053E$-$2&707\p071E$-$2&\nodata      &281\p014E$-$1&364\p019E$-$1&188\p009E$-$1&272\p041E$-$2&112\p022E$-$1\\
NGC5055\c&364\p036E$-$1&663\p067E$-$1&\nodata      &108\p005E$+$1&187\p009E$+$1&254\p012E$+$1&498\p050E$+$1&574\p040E$+$1\\
NGC5229  &182\p018E$-$2&251\p025E$-$2&536\p031E$-$2&122\p006E$-$1&184\p009E$-$1&117\p006E$-$1&444\p115E$-$2&131\p016E$-$1\\
NGC5273  &223\p024E$-$3&104\p015E$-$2&155\p008E$-$1&648\p032E$-$1&127\p006E$+$0&124\p006E$+$0&416\p054E$-$1&294\p031E$+$1\b\\
NGC5523  &402\p040E$-$2&587\p059E$-$2&149\p008E$-$1&395\p020E$-$1&616\p031E$-$1&542\p027E$-$1&749\p075E$-$1&136\p015E$+$0\a\\
NGC5608  &321\p032E$-$2&387\p038E$-$2&660\p039E$-$2&147\p007E$-$1&200\p010E$-$1&109\p005E$-$1&295\p042E$-$2&607\p064E$-$2\a\\

\enddata
\tablecomments{Fluxes (in mJy) are derived using $2a_{25} \times 2b_{25}$ elliptical apertures.  The compact table entry format TUV$\pm$WXYEZ implies (T.UV$\pm$W.XY)$\times10^{\rm Z}$.  All fluxes were corrected for Galactic extinction \citep{schlafly11} assuming $A_V/E(B-V)\approx3.1$ and the reddening curve of \cite{draine03}.  The uncertainties include both statistical and systematic effects.}
\tablenotetext{a} {{\it WISE} 22\m}
\tablenotetext{b} {{\it Herschel} 70\m}
\tablenotetext{c} {Ultraviolet emission extends beyond the aperture \citep{thilker07}.}
\label{tab:fluxes}
\end{deluxetable}



\end{document}